\documentclass[twocolumn]{aastex701}
\usepackage{svg}
\usepackage{pgfplots}
\usepackage{pgfplotstable}
\usepackage{pgf}
\usepackage{import}
\usepackage{adjustbox}
\usepackage{comment}
\usepackage{amsmath}
\usepackage{amssymb}
\usepackage{threeparttable}
\usepackage{booktabs}
\usepackage{subcaption}

\newcommand{\noopsort}[1]{}

\shorttitle{A Partially Stripped Type IIP Supernova}
\shortauthors{Martas et al.}

\received{July 25, 2025}
\revised{August 29, 2025}
\accepted{September 3, 2025}
\submitjournal{ApJ}

\begin{document}

\title{SN\,2019hnl: A Type IIP Supernova with a Partially Stripped, Low Mass Progenitor}

\author[0009-0001-3106-0917]{Aidan Martas}
\affiliation{Department of Astronomy, University of California, Berkeley, CA 94720-3411, USA}
\email[show]{aidmart@berkeley.edu}

\author[0000-0001-8818-0795]{Stefano Valenti}
\affiliation{Department of Physics and Astronomy, University of California, 1 Shields Avenue, Davis, CA 95616-5270, USA}
\email{stfn.valenti@gmail.com}

\author[0000-0002-7352-7845]{Aravind P. Ravi}
\affiliation{Department of Physics and Astronomy, University of California, 1 Shields Avenue, Davis, CA 95616-5270, USA} 
\email{aravind.pazhayathravi@mavs.uta.edu}

\author[0000-0002-7937-6371]{Yize Dong}
\affiliation{Center for Astrophysics \textbar{} Harvard \& Smithsonian, 60 Garden Street, Cambridge, MA 02138-1516, USA}
\email{yize.dong@outlook.com}

\author[0000-0002-4924-444X]{K. Azalee Bostroem}
\altaffiliation{LSST-DA Catalyst Fellow}
\affiliation{Steward Observatory, University of Arizona, 933 North Cherry Avenue, Tucson, AZ 85721-0065, USA}
\email{bostroem@arizona.edu}

\author[0000-0002-0744-0047]{Jeniveve Pearson}
\affiliation{Steward Observatory, University of Arizona, 933 North Cherry Avenue, Tucson, AZ 85721-0065, USA}
\email{jenivevepearson@arizona.edu}

\author[0000-0002-4022-1874]{Manisha Shrestha}
\affil{Steward Observatory, University of Arizona, 933 North Cherry Avenue, Tucson, AZ 85721-0065, USA}
\email{manisha.shrestha82@gmail.com}

\author[0000-0003-0123-0062]{Jennifer E.\ Andrews}
\affiliation{Gemini Observatory, 670 North A`ohoku Place, Hilo, HI 96720-2700, USA}
\email{jandrews@gemini.edu}

\author[0000-0003-4102-380X]{David J.\ Sand}
\affiliation{Steward Observatory, University of Arizona, 933 North Cherry Avenue, Tucson, AZ 85721-0065, USA}
\email{dave.j.sand@gmail.com}

\author[0000-0002-0832-2974]{Griffin Hosseinzadeh}
\affiliation{Department of Astronomy \& Astrophysics, University of California, San Diego, 9500 Gilman Drive, MC 0424, La Jolla, CA 92093-0424, USA}
\email{ghosseinzadeh@ucsd.edu}

\author[0000-0001-9589-3793]{Michael Lundquist}
\affiliation{W.~M.~Keck Observatory, 65-1120 M\=amalahoa Highway, Kamuela, HI 96743-8431, USA}
\email{lund0946@gmail.com}

\author[0000-0003-2744-4755]{Emily Hoang}
\affil{Department of Physics and Astronomy, University of California, 1 Shields Avenue, Davis, CA 95616-5270, USA}
\email{emthoang@ucdavis.edu}

\author[0009-0008-9693-4348]{Darshana Mehta}
\affiliation{Department of Physics and Astronomy, University of California, 1 Shields Avenue, Davis, CA 95616-5270, USA}
\email{ddmehta@ucdavis.edu}

\author[0000-0002-7015-3446]{Nicol\'as Meza Retamal}
\affiliation{Department of Physics and Astronomy, University of California, 1 Shields Avenue, Davis, CA 95616-5270, USA}
\email{nemezare@ucdavis.edu}

\author[0000-0001-8738-6011]{Saurabh W.\ Jha}\affiliation{Department of Physics and Astronomy, Rutgers, the State University of New Jersey, 136 Frelinghuysen Road, Piscataway, NJ 08854-8019, USA}
\email{saurabh@physics.rutgers.edu}

\author[0000-0003-0549-3281]{Daryl Janzen}
\affiliation{Department of Physics \& Engineering Physics, University of Saskatchewan, 116 Science Place, Saskatoon, SK S7N 5E2, Canada}
\email{daryl.janzen@usask.ca}

\author[0000-0003-4253-656X]{D.\ Andrew Howell}
\affiliation{Las Cumbres Observatory, 6740 Cortona Drive, Suite 102, Goleta, CA 93117-5575, USA}
\affiliation{Department of Physics, University of California, Santa Barbara, CA 93106-9530, USA}
\email{ahowell@lco.global}

\author[0000-0001-5807-7893]{Curtis McCully}
\affiliation{Las Cumbres Observatory, 6740 Cortona Drive, Suite 102, Goleta, CA 93117-5575, USA}
\email{curtismccully@gmail.com}

\author[0000-0002-1125-9187]{Daichi Hiramatsu}
\affiliation{Center for Astrophysics \textbar{} Harvard \& Smithsonian, 60 Garden Street, Cambridge, MA 02138-1516, USA}
\affiliation{The NSF AI Institute for Artificial Intelligence and Fundamental Interactions, USA}
\email{daichi.hiramatsu@cfa.harvard.edu}

\author[0000-0002-7472-1279]{Craig Pellegrino}
\affiliation{Goddard Space Flight Center, 8800 Greenbelt Rd, Greenbelt, MD 20771, USA}
\email{craig.m.pellegrino@nasa.gov}

\begin{abstract}

\par We present optical photometry and spectroscopy of SN\,2019hnl.  Discovered within $\sim$26 hr of explosion by the ATLAS survey, SN\,2019hnl is a typical Type IIP supernova with a peak absolute $V$ band magnitude of $-16.7\pm0.1$ mag, a plateau length of $\sim107$ days, and an early decline rate of $0.0086\pm0.0006$ mag (50 days)$^{-1}$.  We use nebular spectroscopy and hydrodynamic modeling with the \textsc{snec}, \textsc{mesa}, and \textsc{stella} codes to infer that the progenitor of SN\,2019hnl was a $M_\text{ZAMS}\sim11M_\odot$ red supergiant which produced $0.047\pm0.007M_\odot$ of $^{56}$Ni in the explosion.  As a part of our hydrodynamic modeling, we reduced hydrogen envelope mass by scaling the mass loss within the ``Dutch" wind scheme to fit our light curve, showing that the progenitor of a relatively typical Type IIP SN may experience partial stripping during their evolution and establish massive ($\sim0.2M_\odot$) CSM environments prior to core collapse.

\end{abstract}


\section{Introduction} \label{sec:intro}

\par Massive stars $\gtrsim8M_\odot$ evolve quickly and end their lifespans in explosive core-collapse supernovae (CCSNe).  Type II supernovae (SNe\,II), the most commonly observed CCSNe \citep{Li2011,Smith2011,Shivvers2017}, display hydrogen in their spectra and have great diversity in photometric and spectral evolution. Historically divided into SNe\,IIP, SNe\,IIL, and SNe\,IIb based upon their photometric and spectroscopic evolution \citep{Patat1994, Arcavi2012, Faran14_IIL}, SNe\,II can maintain relatively constant brightness for $\sim100$ days during the post-peak hydrogen recombination phase (IIP), undergo a linear decline from peak brightness (IIL), or gradually eliminate hydrogen from their spectra (IIb).  With the collection of larger samples of SNe\,II, it has become evident that SNe IIP and IIL are likely a continuous class of objects \citep[e.g.,][]{Anderson2014, Valenti2016}.

\par While red supergiants (RSGs) are known to be the progenitors of SNe\,IIP and SNe\,IIL \citep{Smartt2015}, the mass range of RSGs ending their life as CCSNe is still uncertain.  SNe\,II progenitors surrounded by a greater mass of circumstellar material (CSM) become more luminous at their peak and exhibit a more rapid, linear decline, yielding SNe\,IIL (linear) \citep{Morozova2017, Morozova2018,Hiramatsu2021}.  The existence of a continuum between SNe\,IIP and SNe\,IIL is supported by optical spectra \citep{Valenti2015, Valenti2016}, while near-infrared spectra suggest a discontinuity - albeit with a smaller sample size \citep{Davis2019}.  At exceptionally high CSM densities, SNe\,II can exhibit narrow hydrogen emission lines due to ejecta-CSM interaction ionizing the unshocked CSM, yielding the SNe\,IIn (narrow) subclass \citep{IIN}.  Prior to explosion, some massive stars' outer hydrogen and helium envelopes are stripped away, creating SNe\,IIb, which initially show hydrogen in their spectra, only for it to weaken or disappear at later times \citep{SNe_IIb}.  A recent analysis of SNe\,II based upon hydrodynamic modeling \citep{Fang_2025} has even suggested that envelope stripping may be a feature common to and instrumental in the photometric and spectroscopic diversity within SNe\,II.

\par While stripping may be frequent, understanding of the stripping mechanisms is limited.  Strong stellar winds (``superwinds") have been proposed as a source of mass ejection \citep{winds_are_real}, though their validity is contentious for the majority of RSGs \citep{winds_arent_real}.  More recently, binary systems have gained traction as a mass loss pathway capable of ejecting the necessary mass to match observations \citep{binary1, binary2, binary3, binary4}.  A dearth of direct progenitor data hinder understanding of the physics behind envelope stripping, delaying definitive conclusions.

\par Over the past two decades, significant progress has been made in the computational modeling of stellar evolution and explosions.  Hydrodynamic models are now being compared with photometric and spectroscopic observations to estimate progenitor metrics, envelope stripping, and CSM geometry \citep{JerkstrandCode, JerkstrandModels, SNEC, Hiramatsu2021}, often resulting in closely matching results.  In this paper, we apply these methods to determine progenitor properties.

\par Here, we present optical photometry and spectroscopy of SN\,2019hnl and apply hydrodynamic modeling to determine progenitor properties.  In Section \ref{sec:observations}, we report the discovery and photometric and spectroscopic observations.  In Section \ref{sec:obs_props}, we report the observational properties of SN\,2019hnl, including its reddening, light curve, and spectroscopic evolution.  In Section \ref{sec:discussion}, we estimate the $^{56}$Ni mass, establish the presence of partial stripping, and place an upper bound on the progenitor's zero age main sequence (ZAMS) mass.  In Sections \ref{sec:snec} and \ref{sec:mesa}, we use hydrodynamic modeling to ascertain progenitor properties.  Finally, we present our conclusions in Section \ref{sec:conclusion}.

\begin{figure}
    \centering
    \includegraphics[width=\linewidth]{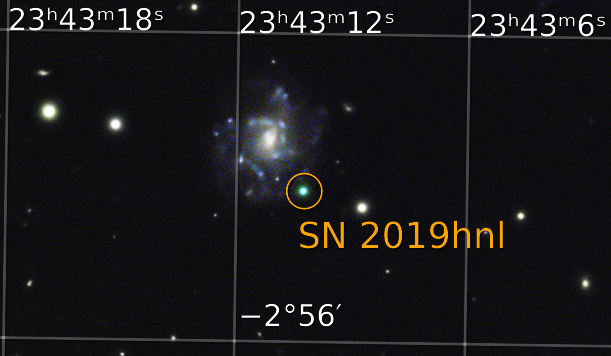}
    \caption{Stacked RGB image of SN\,2019hnl using the $BVg$ bandpasses.  SN\,2019hnl is in the outskirts of the host, aligning with our low host reddening discussed in Section \ref{sec:reddening}.}
    \label{fig:findingchart}
\end{figure}

\section{Observations} \label{sec:observations}

\subsection{Photometry} \label{sec:photometry}
\par SN\,2019hnl was discovered at R.A. 23$^{\text{h}}$ 43$^{\text{m}}$ 10.263$^{\text{s}}$, Dec. -2$^{\text{h}}$ 56$^{\text{m}}$ 58.64$^{\text{s}}$ (J2000) on 2019-06-14 13:39:21 (UT) at $18.6$ mag ($o$) by the Asteroid Terrestrial-impact Last Alert System (ATLAS) program \citep{Discovery} in the spiral galaxy RASSCALS SS2b312.003.  Two $5\sigma$ nondetections near explosion were reported by the Zwicky Transient Facility \citep[ZTF;][]{Bellm19, Graham19, Masci23} through their forced photometry service in the $g$ band at $20.2$ mag and $20.0$ mag 4.1 and 1.1 days before the ATLAS discovery on 2019-06-13 11:16:25, respectively.  We ran ATLAS forced photometry at the location of SN\,2019hnl, but no further limits were found as the field had not been observed between the latest ZTF nondetection and discovery.  We adopt the later nondetection as the explosion time $t_0$.  SN\,2019hnl was classified as a SN\,II 5 days after discovery \citep{Classification} by the Global Supernova Project \citep[GSP;][]{Howell19}.

\par
\begin{figure*}
    \centering
    \includegraphics[width=\linewidth]{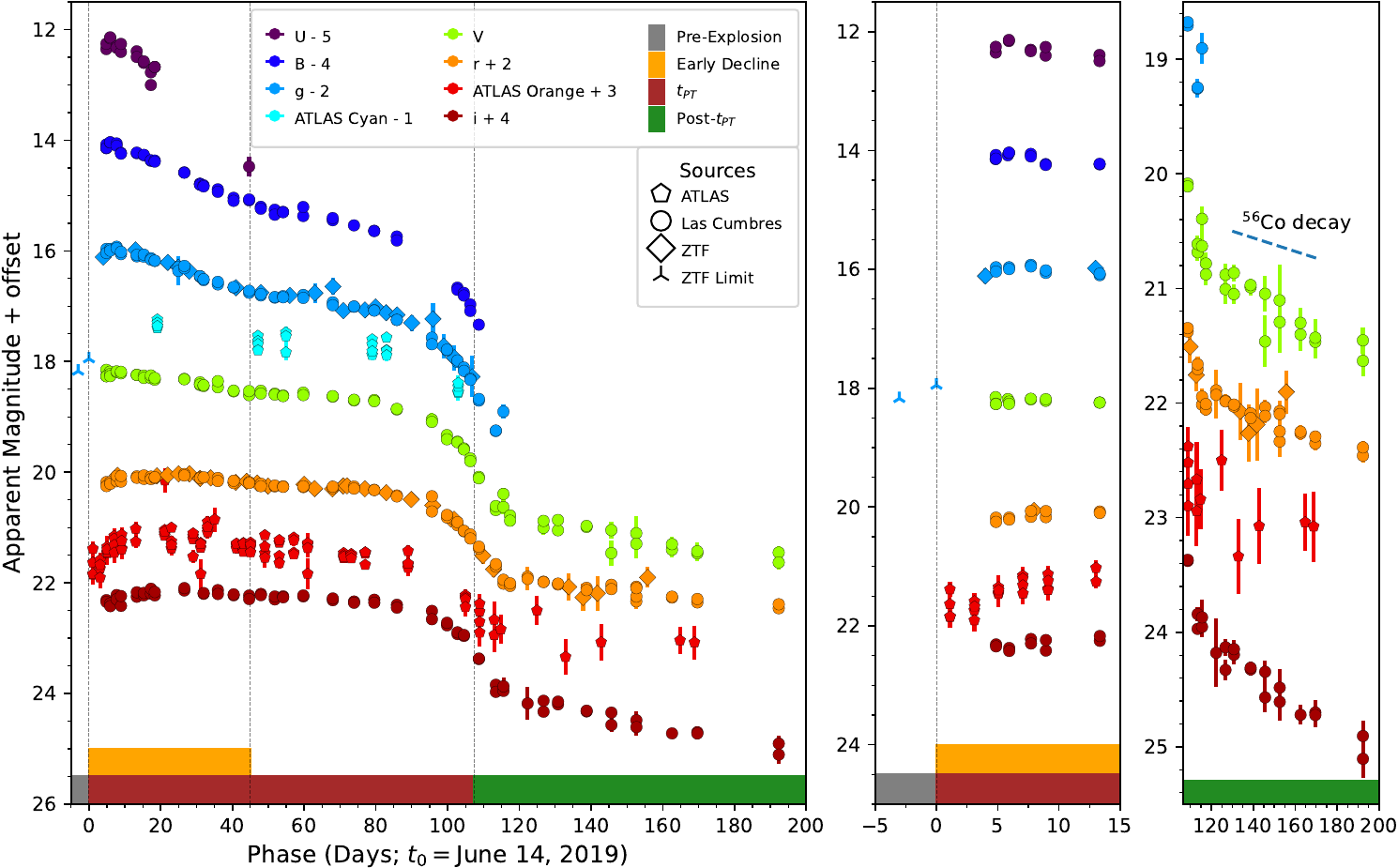}
    \caption{\emph{Left panel:} Multiband light curves of SN\,2019hnl from Las Cumbres, ZTF, and ATLAS.  Bars on the base of each panel denote the phases of photometric evolution in accordance with \citet{Valenti2016}.  The last nondetection was one day before the ATLAS detection in the $o$ filter.  \emph{Center panel:} Zoomed-in view from $t_0-5$ days to $t_0+15$ days.  \emph{Right panel:} Zoomed-in view from $t_\text{PT}$ onwards.}
    \label{fig:lightcurve}
\end{figure*}

 The GSP triggered photometric observations from the Las Cumbres Observatory 1-m telescope network \citep{Brown2013} 4 days following discovery.  Photometric data were reduced with the \texttt{lcogtsnpipe} pipeline \citep{Valenti2016}.  Data for the $gri$ filters were calibrated using stars in the APASS catalog \citep{apass}, while $UBV$ data were calibrated using standard stars from the Landolt catalog \citep{landolt} observed with the same telescope on the same night.  Given the negligible host contamination discussed in Section \ref{sec:reddening}, we measured PSF photometry without reference subtraction.  In addition to the Las Cumbres data, we acquired forced photometry from both ATLAS ($o$,$c$) and the ZTF ($g$,$r$).  Photometry points are plotted in Figure \ref{fig:lightcurve}.

\subsection{Spectroscopy} \label{sec:spectroscopy}

\par
\begin{figure*}
    \centering
    \includegraphics[width=\linewidth]{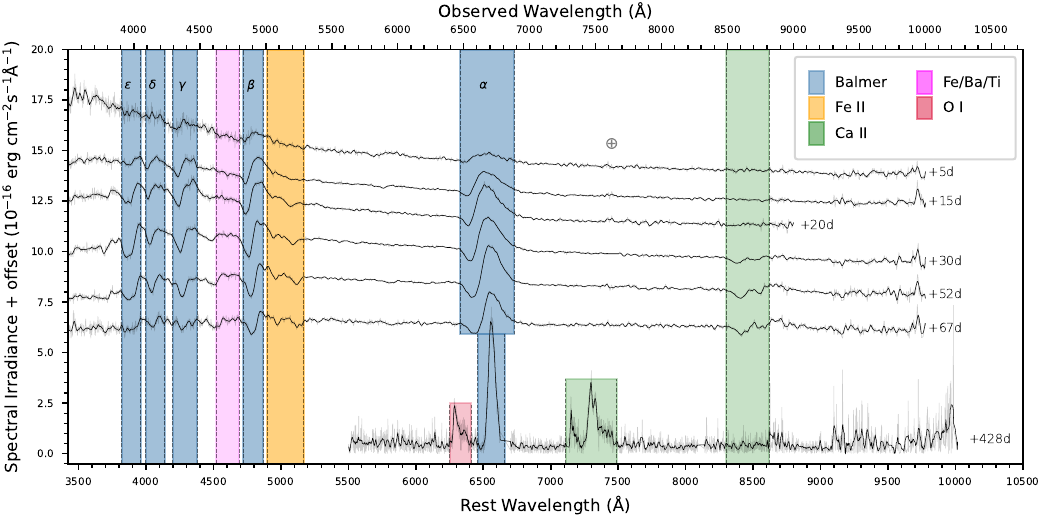}
    \caption{Spectroscopic evolution of SN\,2019hnl in optical wavelengths from $+5$ days to $+428$ days.  Spectra smoothed with a Savitsky-Golay filter are plotted in black over the gray, unsmoothed spectra.}
    \label{fig:specplot}
\end{figure*}

 We collected six spectra from the 2m FLOYDS spectrograph \citep{Brown2013} through the GSP between 5 and 67 days post-explosion and one nebular spectrum from the Low-Resolution Imaging Spectrometer \citep[LRIS;][]{LRIS} on the Keck I telescope at 428 days post-explosion.  FLOYDS spectra were taken with a $2"\times30"$ slit aligned with the parallactic angle and reduced using the FLOYDS reduction pipeline \citep{Valenti2014}.  After flux calibration, all spectra were scaled to $i$-band photometry at the same epoch.  In the case of the nebular spectrum, we scaled to $i$-band photometry linearly extrapolated from the radioactive tail ($m_i = 23.96$ mag) as no photometry was available at the epoch of the spectrum.  All spectra are plotted in Figure \ref{fig:specplot} and the log of spectra is shown in Table \ref{tab:speclog}.

\begin{deluxetable*}{cccccc}
\tablenum{1}
\tablecaption{SN\,2019hnl Spectra \label{tab:speclog}}
\tablewidth{0pt}
\tablehead{
\colhead{UTC Date \& Time (hh:mm)} & \colhead{MJD (Days)} & \colhead{Phase (Days)} &
\colhead{Telescope} & \colhead{Instrument} & \colhead{Wavelength Coverage (\AA)} }
\startdata 
    2019-06-19 12:54 & 58,653.537 & 5 & FTN & FLOYDS & 3422 -- 9777 \\
    2019-06-29 13:44 & 58,663.572 & 15 & FTN & FLOYDS & 3422 -- 9777 \\
    2019-07-04 12:59 & 58,668.541 & 20 & FTN & FLOYDS & 3422 -- 8800 \\
    2019-07-14 13:36 & 58,678.567 & 30 & FTN & FLOYDS & 3422 -- 9777 \\
    2019-08-05 10:12 & 58,700.425 & 52 & FTN & FLOYDS & 3422 -- 9777 \\
    2019-08-20 11:28 & 58,715.478 & 67 & FTN & FLOYDS & 3422 -- 9777 \\
    2020-08-15 12:48 & 59,076.534 & 428 & Keck I & LRIS & 3422 -- 10000 \\
\enddata{}
\end{deluxetable*}

\section{Observational Properties} \label{sec:obs_props}

\subsection{Reddening} \label{sec:reddening}

\par
\begin{figure}
    \centering
    \includegraphics[width=\linewidth]{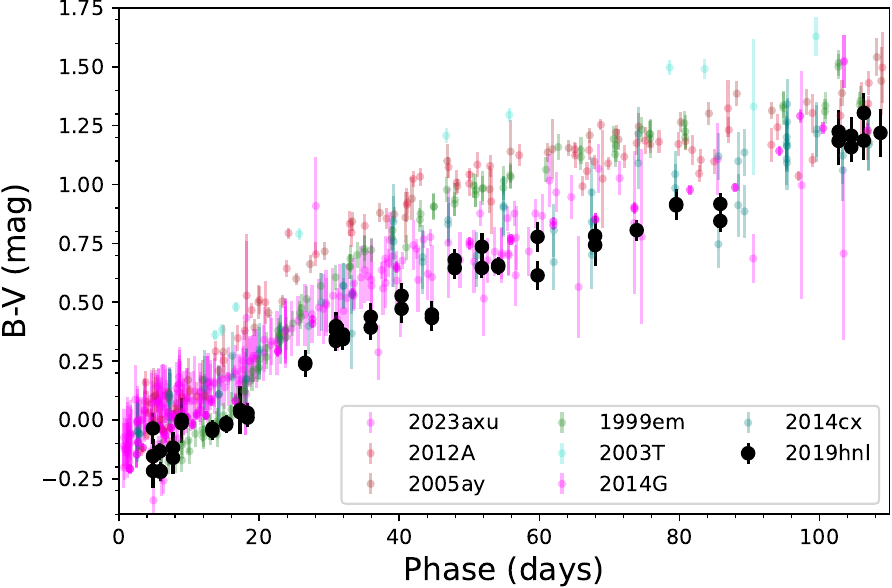}
    \caption{Color evolution of SN\,2019hnl; photometrically similar SNe\,IIP SN\,1999em \citep{Faran14_IIP}, SN\,2005ay \citep{05ay_2,05ay_1}, SN\,2012A \citep{deJaeger2019}, and SN\,2023axu \citep{Shrestha_axu}; and low-reddening SNe\,II SN\,2003T \citep{03T_2, 03T_1}, SN\,2014G \citep{deJaeger2019}, and SN\,2014cx \citep{Valenti2016}.  All points are corrected for both MW and host reddening, when applicable.  SN\,2019hnl's $B-V$ evolution is similar to, but on the bluer end of these other SN\,II.}
    \label{fig:bvplot}
\end{figure}

 The Na \textsc{id} $\lambda\lambda5890,5896$ doublet from Milky Way (MW) and SN host extinction are not clearly detected, suggesting both low MW and host galactic reddening \citep{Reddening2, Reddening1}.  We find $3\sigma$ upper limits of $0.33$\AA \text{ } and $0.49$\AA \text{ }for the pseudo-equivalent-widths (pEWs) of the MW and host Na \textsc{id} lines, respectively, implying an upper limit of $E(B-V)_\text{host} \sim 0.04$ mag.  Despite the absence of MW Na \textsc{id} in the spectra, we still use the small line-of-sight $E(B-V)_\text{MW}=0.0293\pm 0.0009$ mag reported in the dust maps of \cite{MWReddening}, while we assume no reddening for the host.  This value is consistent with our upper limits and we adopt $E(B-V)_\text{tot} = E(B-V)_\text{MW} = 0.0293\pm 0.0009$ mag with $R_V = 3.1$ \citep{Cardelli} for this paper.  However, the host reddening relations from \cite{Reddening1} are known to underestimate the uncertainty in reddening \citep{IS_Absorption}.  Therefore, the assumption of zero host reddening should be treated with some caution.  As a sanity check, we plot the dereddened $B-V$ color for SN\,2019hnl with photometrically similar SNe\,IIP in Figure \ref{fig:bvplot}.  SN\,2019hnl appears slightly bluer than the comparison SNe during plateau, but reaches a typical color following plateau at $>100$d, also supporting the low extinction towards SN\,2019hnl.

\subsection{Distance} \label{sec:distance}

\par We applied the expanding photosphere method (EPM) to SN\,2019hnl using our spectral data, though our result contained significant variance across filter combinations.  Since there are no other independent distance measurements for the host galaxy, we fix the Hubble parameter $H_0=69.3 \text{ km s}^{-1}\text{ Mpc}^{-1}$ \citep{9yWMAP} and adopt a derived Hubble flow distance of $97.7$ Mpc based on the host redshift $z=0.023$ \citep{HostRedshift}.  We detail our EPM attempt in Appendix \ref{appendix:epm}.

\subsection{Photometric Evolution} \label{sec:phot_evo}

\par Multiband photometry is presented in Figure \ref{fig:lightcurve}.  The $V$ band shows a rise in brightness to a maximum of $M_V=-16.7\pm0.1$ mag $\sim7$ days post-explosion, while the $g$-band light curve shows a more pronounced maximum of $M_g=-17.0\pm0.2$ about $6.5$ days post-explosion.  The brightness then remains nearly constant until $\sim100$ days post-explosion, primarily powered by the optically thick hydrogen recombination front.  Following this plateau phase, the brightness decreases and eases into a linear decline phase powered by the $^{56}\text{Ni}\rightarrow^{56}\text{Co}\rightarrow^{56}\text{Fe}$ decay chain.

\par
\begin{figure}
    \centering
    \includegraphics[width=\linewidth]{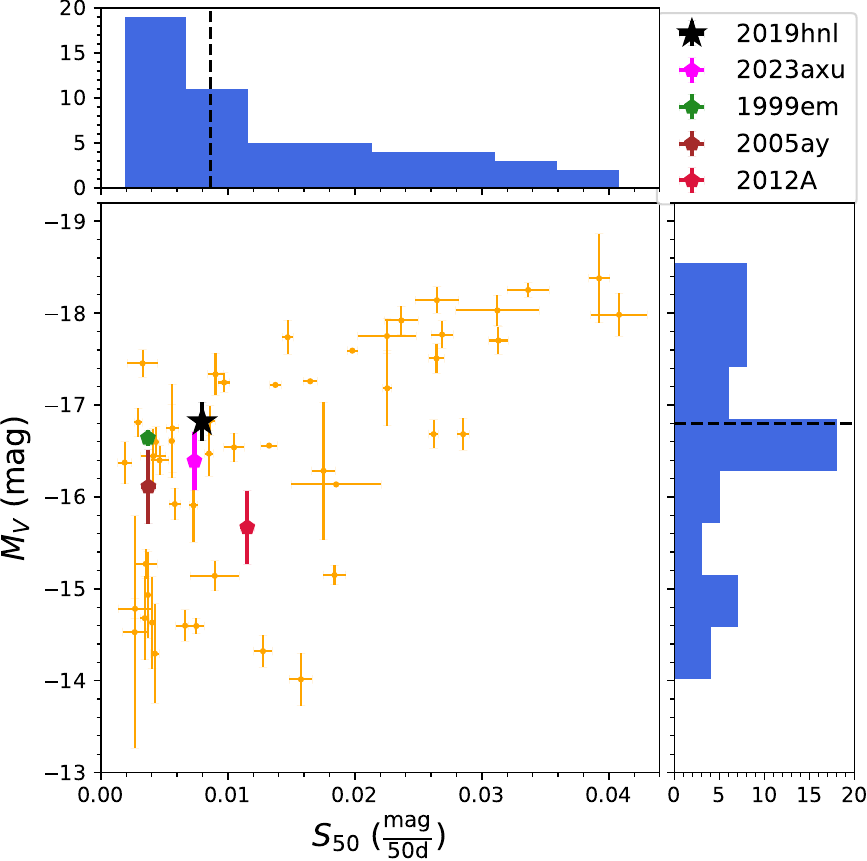}
    \includegraphics[width=\linewidth]{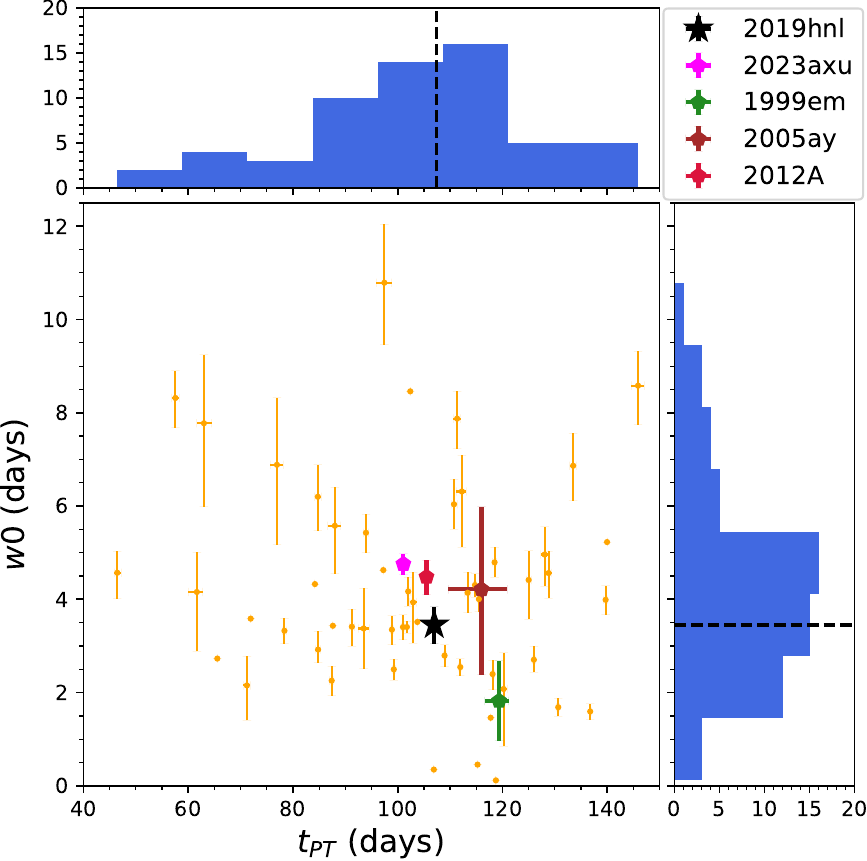}
    \caption{\emph{Upper panel:} SN\,2019hnl and comparison SN\,II from SNDAVIS in $S_{50}-M_V$ parameter space.  \emph{Lower panel:} SN\,2019hnl, and comparison SN\,II from SNDAVIS in $t_\text{PT}-w0$ parameter space.  SN\,2019hnl's position in both parameter spaces is consistent with the general SNe\,II population.}
    \label{fig:s50mv_s1tpt}
\end{figure}

\begin{figure*}
    \centering
    \includegraphics[width=\linewidth]{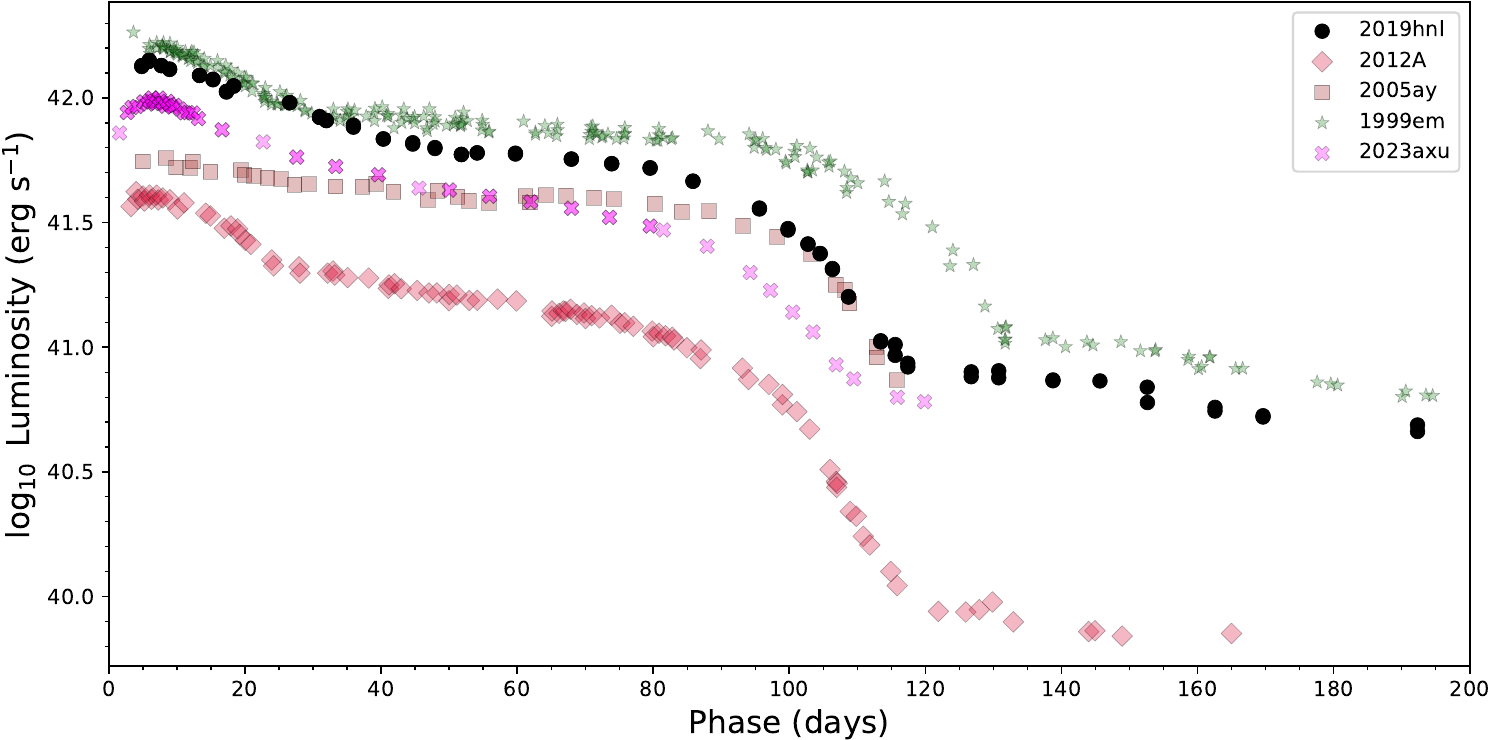}
    \caption{Pseudobolometric evolution comparison of SN\,2019hnl and other SNe\,IIP from the SNDAVIS database, all synthesized from $UBVgri$ photometry.  SN\,1999em displays a similar evolution up to the end of early decline, while SN\,2023axu evolves analogously but almost uniformly dimmer by $\sim0.4$ mag.}
    \label{fig:bolocomp}
\end{figure*}

 Following maximum brightness, SNe\,II can evolve along a wide range of photometric tracks.  To locate SN\,2019hnl in the SNe\,II continuum, we measured the rate of change of $V$-band brightness per 50 days $S_{50}$ in accordance with definitions in \cite{Valenti2016}.  We find $S_{50}=0.0086\pm0.0006 \text{ mag (50 days)}^{-1}$.  When plotted with other SNe\,II with similar $t_\text{PT}$, $S_{50}$, or $M_V$ (see Figure \ref{fig:bolocomp}) from SNDAVIS\footnote{\href{https://dark.physics.ucdavis.edu/sndavis}{https://dark.physics.ucdavis.edu/sndavis}} \citep{Faran14_IIP,Anderson2014,Valenti2016,deJaeger2019,Anderson2024, Shrestha_axu} in Figure \ref{fig:s50mv_s1tpt}, SN\,2019hnl lies firmly in the typical SNe\,IIP part of the $S_{50}$-$M_V$ parameter space.

\par Following plateau, the subsequent dimming can be modeled empirically as a Fermi-Dirac function \citep{Valenti2016}

\begin{align}
    y(t) &= -\frac{a0}{1+e^{t-t_\text{PT}/w0}} + p0\cdot(t-t_\text{PT})+m0
    \label{eqn:fermidir}
\end{align}

where $a0$ represents the dimming depth and $w0$ inversely represents the slope of the light curve following $t_\text{PT}$ but before the $^{56}\text{Ni}$ tail.  We determined $t_\text{PT}=107.4\pm0.4\text{ days}$, $a0=1.48\pm0.04 \text{ mag}$, and $w0=3.45^{+0.42}_{-0.38}\text{ days}$ by fitting the $V$-band photometry with Equation \ref{eqn:fermidir} using the MCMC sampling Python package \texttt{emcee} \citep{emcee}.  We plot $w0$ versus $t_\text{PT}$ in Figure \ref{fig:s50mv_s1tpt}, showing that SN\,2019hnl is located solidly within typical SNe\,II parameter space.

\par We constructed a pseudobolometric optical light curve for SN\,2019hnl with our $UBVgri$ photometry by integrating each filter's flux with Simpson's rule in accordance with \cite{LCSynthesis}.  In Figure \ref{fig:bolocomp}, we compare the pseudobolometric optical light curves of several typical SNe\,IIP to that of SN\,2019hnl.  The geometry of SN\,2019hnl's pseudobolometric evolution is most similar to that of SN\,2023axu \citep{Shrestha_axu}, which is nearly uniformly dimmer by $\sim0.4$ magnitudes; and SN\,1999em \citep{Faran14_IIP}, which has a slightly longer plateau.  In conclusion, the light curve of SN\,2019hnl is typical for moderately luminous SNe\,IIP.

\subsection{Spectral Evolution} \label{sec:spec_evo}

\par
\begin{figure}
    \centering
    \includegraphics[width=\linewidth]{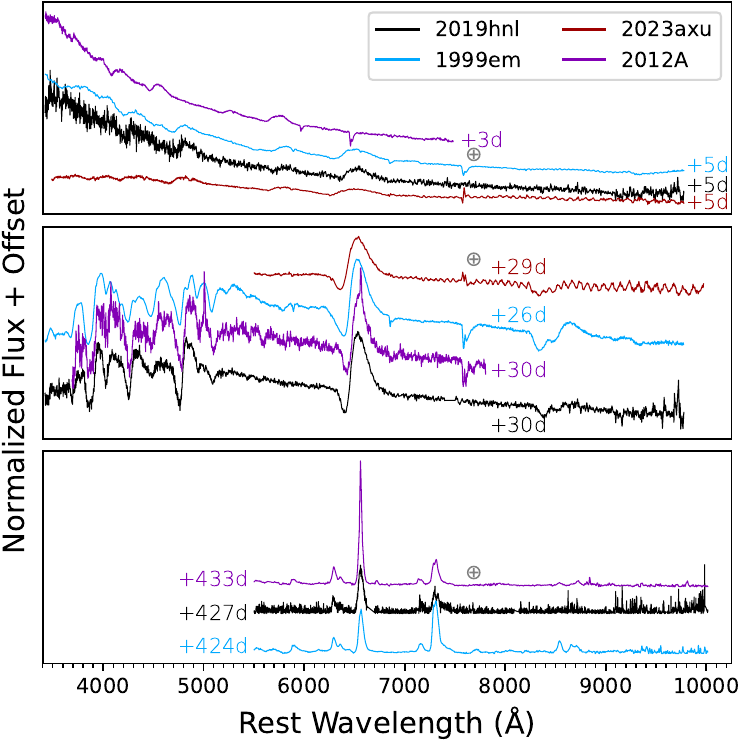}
    \caption{Spectral comparison of SN\,2019hnl with similar-phase spectra for SN\,1999em \citep{99em_spec1, 99em_spec2}, SN\,2023axu \citep{Shrestha_axu}, and SN\,2012A \citep{12A_spec1, 99em_spec2}.  \emph{Upper panel:} Early spectra at $\sim5$d.  The pattern on the red side of the $+29$d SN\,2023axu spectrum is due to fringing.  \emph{Center panel:} Plateau spectra during recombination at $\sim30$d.  \emph{Lower panel:} Nebular spectra at $\sim427$d.}
    \label{fig:speccomp}
\end{figure}

 The spectral evolution of SN\,2019hnl is typical of SNe\,IIP; H$\alpha$ and H$\beta$ become visible at early times and continue to strengthen as time passes, while \ion{Fe}{2} becomes visible $\sim20$ days post-explosion.  The \ion{Ca}{2} line remains hidden until $\sim30$ days post-explosion and continues to strengthen over time within our plateau spectra.  Given the similar photometric evolution of SN\,1999em, SN\,2012A, SN\,2019hnl, and SN\,2023axu, we compare one spectrum for each SN at early, plateau, and nebular phases in Figure \ref{fig:speccomp}.  All nebular spectra are flux calibrated to contemporaneous photometry.

\par
\begin{figure}
    \centering
    \includegraphics[width=\linewidth]{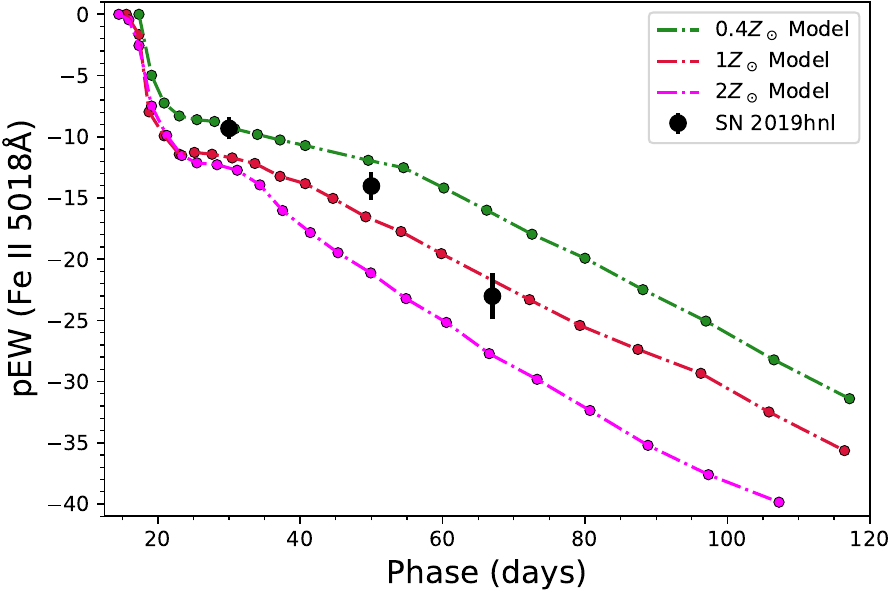}
    \caption{pEWs of the Fe $\lambda5018$ line for SN\,2019hnl and the models discussed in \cite{Z_Comp}.  SN\,2019hnl falls between the $0.4Z_\odot$ and $Z_\odot$ models at early times, but migrates to between the $Z_\odot$ and $2Z_\odot$ models at $+67$ days.}
    \label{fig:z_comp}
\end{figure}

 Of note are the weak \ion{Fe}{2} lines at $+30$ days for SN\,2019hnl, introducing the possibility of a low-metallicity progenitor.  \cite{Z_Comp} found that model \ion{Fe}{2} lines at 4923, 5018, and 5169 \AA\text{ }intensified as metallicity increased and weakened as metallicity decreased for photospheric spectra, finding that SN\,2007il, SN\,2005J, and SN\,2008ag matched $0.4Z_\odot$, $1Z_\odot$, and $2Z_\odot$ models, respectively.  In Figure \ref{fig:z_comp}, we compare the pEWs of the \ion{Fe}{2} lines in SN\,2019hnl to those from models.

\par The \ion{Fe}{2} lines of SN\,2019hnl show pEWs most consistent with those of subsolar metallicity models for early times, but trends towards the supersolar metallcity model at late times.  The model metallicities appear to be following a slower temporal evolution than that of SN\,2019hnl, which could be caused by a more rapid temperature drop.  We conclude that the pEW evolution is unlikely to be governed exclusively by metallicity and likely is affected by other explosion properties.

\par
\begin{figure}
    \centering
    \includegraphics[width=\linewidth]{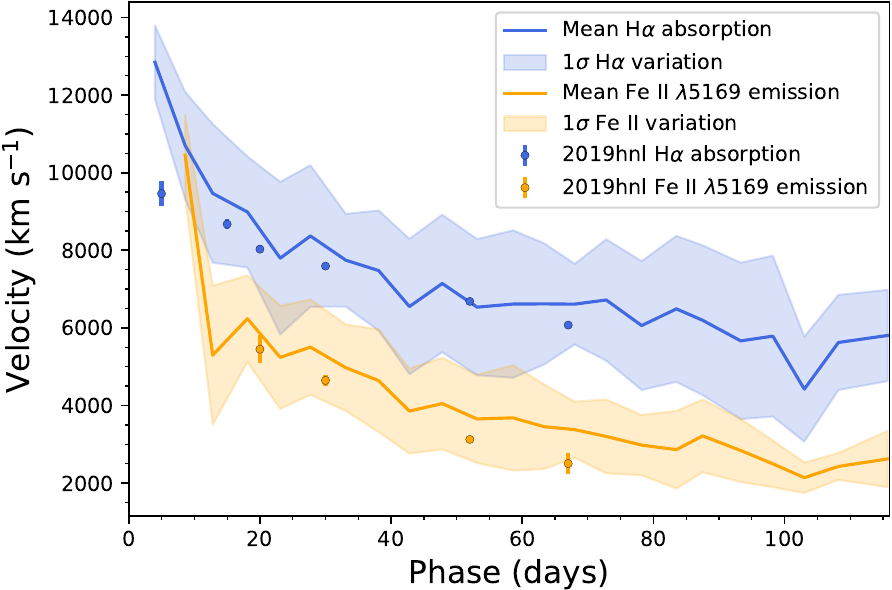}
    \caption{Velocity evolution of the \ion{Fe}{2} and H$\alpha$ lines in SN\,2019hnl (points) compared to a sample of SNe\,II (bands) from \cite{GutierrezVelos}.}
    \label{fig:linevelo}
\end{figure}

 We measure the Fe and H$\alpha$ velocities from the P Cygni minima and compare with the average velocities of 122 SNe\,II measured by \cite{GutierrezVelos} in Figure \ref{fig:linevelo}.  The velocity of hydrogen lines is expected to be greater than that of Fe lines because the hydrogen lines form in the external layers of the ejecta, while Fe lines form in the inner layers of the ejecta.  Both line velocities are consistent with a slightly below-average velocity.

\section{Discussion} \label{sec:discussion}

\subsection{Nickel Mass} \label{sec:nickel}

\par The synthesized nickel mass of SNe\,IIP can be estimated based upon the luminosity decline during the nebular phase, which is powered by the decay chain $^{56}\text{Co}\rightarrow^{56}\text{Fe}$.  This decay chain produces $\gamma$-rays, which are reprocessed into the optical spectrum by the SN ejecta.  If the ejecta completely traps the $\gamma$-rays, one can compare the pseudobolometric light curve with that of SN\,1987A to estimate the synthesized $^{56}$Ni mass, presuming the two SNe share the same spectral energy distribution.  SN\,2019hnl's brightness decays similarly ($\sim0.0125$ mag day$^{-1}$) to the expected rate for complete $\gamma$-ray trapping ($\sim0.0098$ mag day$^{-1}$) and we therefore assume complete trapping.  Equation \ref{eqn:nickel1} shows the relation between luminosity and synthesized $^{56}$Ni mass $M_{^{56}\text{Ni}}$ presented in \cite{Nickel1}

\begin{align}
    M_{^{56}\text{Ni}} &= 0.075M_\odot\cdot\frac{L_\text{SN}(t)}{L_\text{87A}(t)}
    \label{eqn:nickel1}
\end{align}

where $M_{^{56}\text{Ni}}$ is the synthesized $^{56}$Ni mass, while $L_\text{SN}(t)$ and $L_\text{87A}(t)$ are the pseudobolometric luminosities of the two SNe at time $t$.

\par
\begin{figure}
    \centering
    \includegraphics[width=\linewidth]{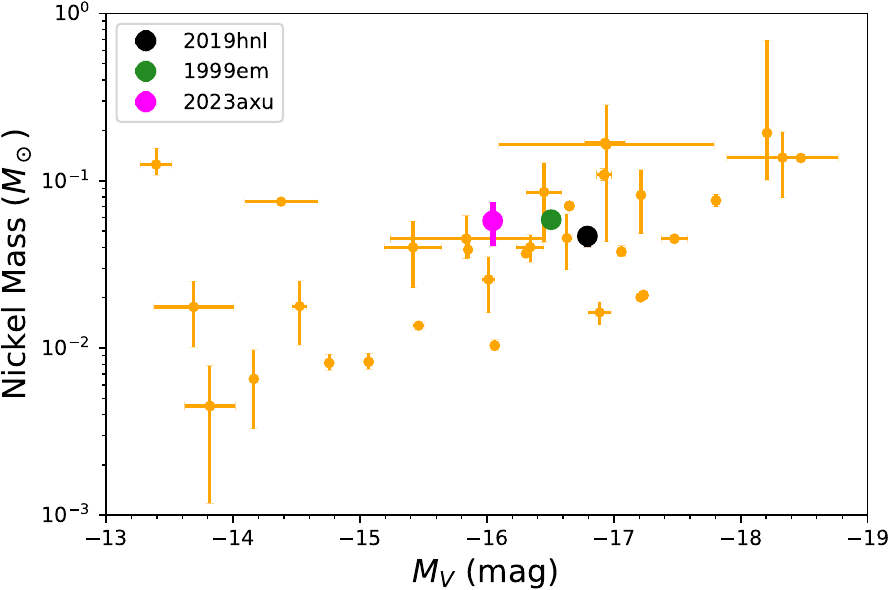}
    \caption{SN\,2019hnl plotted with other SNe\,II from SNDAVIS in $M_V-^{56}$Ni parameter space.  Our estimate for the $^{56}$Ni produced in SN\,2019hnl lies within the range of typical $^{56}$Ni production for SNe\,II and near the photometrically similar SN\,1999em and SN\,2023axu.}
    \label{fig:ni_mv}
\end{figure}

 To determine the $^{56}$Ni mass synthesized in SN\,2019hnl, we compared the pseudobolometric photometry (Section \ref{sec:phot_evo}) in the nebular phase ($t>120$ days) with that of SN 1987A using MCMC sampling to determine $M_{^{56}\text{Ni}} = 0.047\pm0.007M_\odot$.  As a sanity check, we plot the $M_{^{56}\text{Ni}}$ versus $M_V$ for our estimation and for other SNe\,II from the SNDAVIS database in Figure \ref{fig:ni_mv}.  SN\,2019hnl's $^{56}$Ni mass aligns with that of other SNe\,II at similar $M_V$.

\subsection{Progenitor Properties}

\subsubsection{Nebular Spectroscopy} \label{sec:nebular}

\par As the ejecta expands and its density and temperature drop, the ejecta becomes optically thin and the inner ejecta geometry and composition is revealed.  The intensity of the \ion{O}{1} $\lambda\lambda6300,6364$ doublet at these times is believed to be correlated with progenitors' ZAMS mass \citep{JerkstrandModels} as more massive progenitors are expected to synthesize more oxygen over their lifetimes.  \cite{JerkstrandModels} modeled nebular spectra for $12M_\odot$, $15M_\odot$, $19M_\odot$, and $25M_\odot$ progenitors, with stellar evolution and supernova explosion modeled by \textsc{kepler} \citep{KeplerCode} and spectra synthesized with \textsc{sumo} \citep{JerkstrandCode}.  To constrain the progenitor mass, we compare the intensities of the \ion{O}{1} $\lambda\lambda6300,6364$ doublet in our nebular spectrum.

\par
\begin{figure*}
    \centering
    \includegraphics[width=\linewidth]{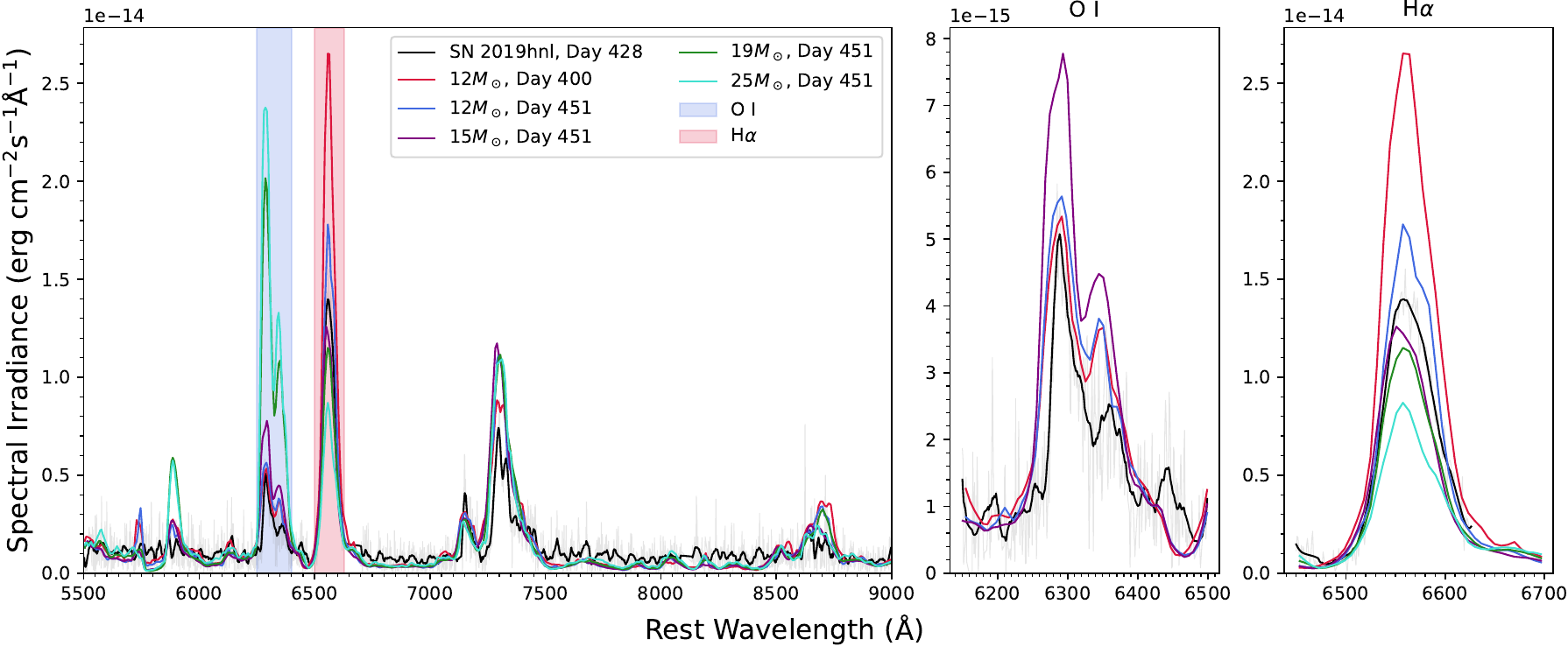}
    \caption{\emph{Left panel:} Nebular spectrum of SN\,2019hnl at $+428$d compared with scaled synthetic spectra from \cite{JerkstrandModels} for varying progenitor masses.  The nebular spectrum smoothed with a Savitsky-Golay filter is plotted above the gray, unsmoothed spectrum.  \emph{Center panel:} Zoomed-in view of the \ion{O}{1} $\lambda\lambda6300,6364$ doublet.  The nearest model corresponds to a ZAMS mass of $12M_\odot$ and has a more intense profile, implying a $\sim12M_\odot$ upper limit to ZAMS mass.  \emph{Right panel:} Zoomed-in view of the H$\alpha$ profile.  The $12M_\odot$ model significantly overestimates the line intensity, implying that hydrogen may have been removed from the progenitor, possibly by envelope stripping.}
    \label{fig:nebspec}
\end{figure*}

 To compare the model spectra with the $t+428$d nebular spectrum, we scale the models' synthetic $i$ band fluxes to the same photometry point linearly extrapolated from the $^{56}$Ni tail.  The comparison is shown in Figure \ref{fig:nebspec}.

\par The higher-mass models significantly overestimate the \ion{O}{1} intensity.  The $12M_\odot$ model yields the closest \ion{O}{1} intensity, but still overestimates by $\sim20-40\%$.  We conclude that the nebular modeling suggests a progenitor mass $\lesssim12M_\odot$.

\par While the \ion{O}{1} intensity is consistent with a low-mass progenitor, the H$\alpha$ intensity is closer to that of a $19M_\odot$ progenitor, though a relationship between progenitor mass and H$\alpha$ luminosity is not necessarily anticipated.  Linearly interpolating between the $t+400$d and $t+451$d $12M_\odot$ model spectra, SN\,2019hnl's $H\alpha$ is less luminous by $\sim60\%$.  This difference may be due to partial stripping of the progenitor's hydrogen envelope, leading to a weaker H$\alpha$ profile in the nebular phase, though \cite{99em_spec2} found that the models overproduce H$\alpha$ emission.  These models use unstripped progenitors and may not reflect properties of partially stripped progenitors, though this is unlikely to be a problem for mass estimation as the modeling of partially-stripped progenitors discussed in Section \ref{sec:mesa} matches these models' low mass estimation.

\subsubsection{Hydrodynamic Modeling: SNEC} \label{sec:snec}

\par We estimated the progenitor mass by comparing SN\,2019hnl's $gri$ light curve with that of hydrodynamic models.  For our first model grid, we used the \textsc{snec} code \citep{SNEC} to explore constraints for progenitor mass, explosion energy, and CSM density and extent.  \textsc{snec} assumes the ejecta has a spherical geometry and that the medium is in local thermodynamic equilibrium (LTE).  Since the LTE assumption is most valid until the nebular phase, we restricted photometry to before $t_\text{PT}$.  We initialized our models with nonrotating, solar metallicity RSGs \citep{Sukhbold2016}.  For CSM models, a $r^{-2}$ density profile was added around the model star with a scaling parameter $\rho_0$ varied to change overall CSM density as expressed in Equation \ref{eqn:csmdensity}, where $\dot{M}$ is the mass loss rate and $v$ the wind velocity.

\begin{align}
    \rho(r) &= \frac{\dot{M}}{4\pi r^2v} = \frac{\rho_0}{r^2}
    \label{eqn:csmdensity}
\end{align}

\par
\begin{figure*}
    \centering
    \begin{minipage}{0.48\textwidth}
        \includegraphics[width=\linewidth]{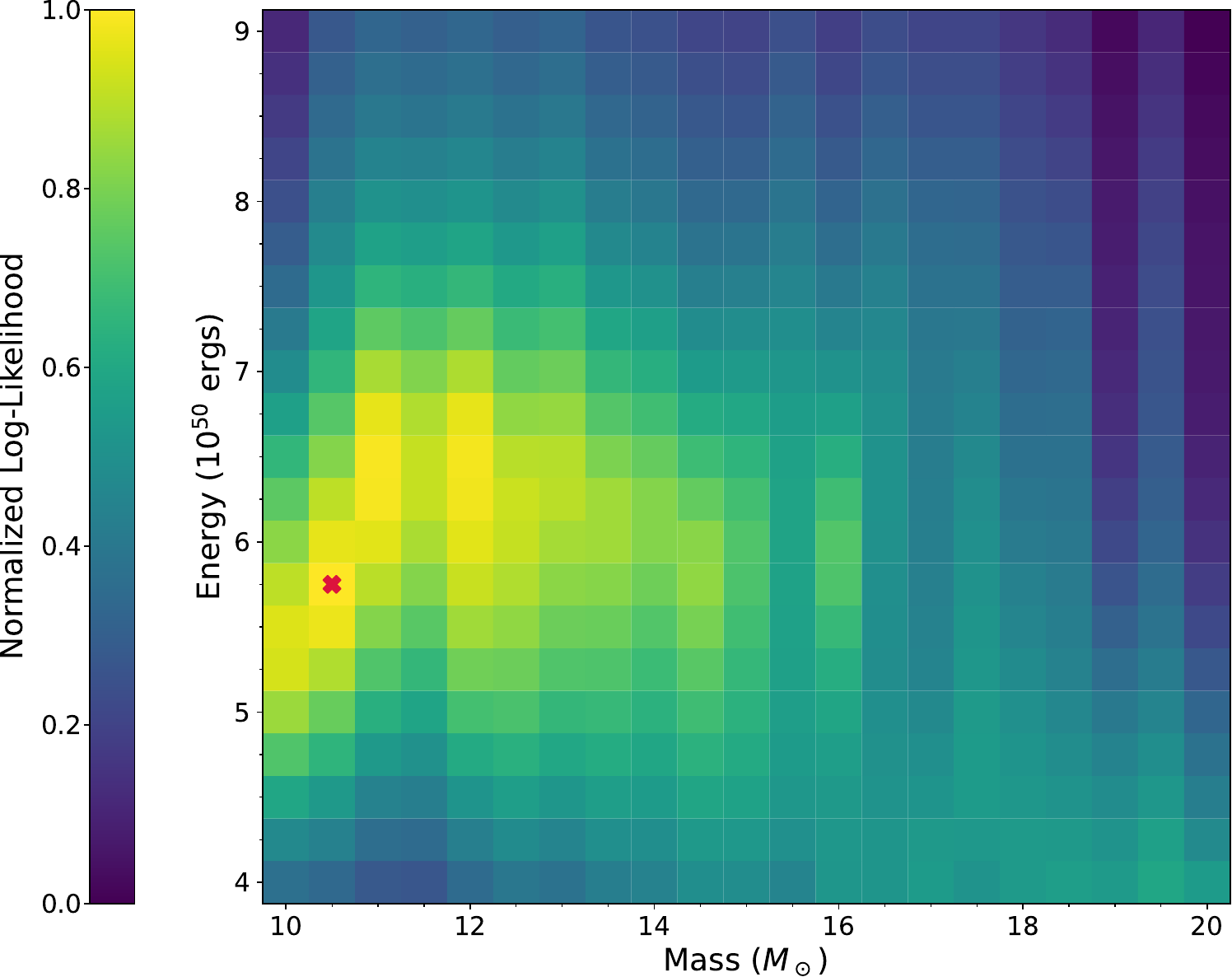}
    \end{minipage}\hfill
    \begin{minipage}{0.48\textwidth}
    \includegraphics[width=\linewidth]{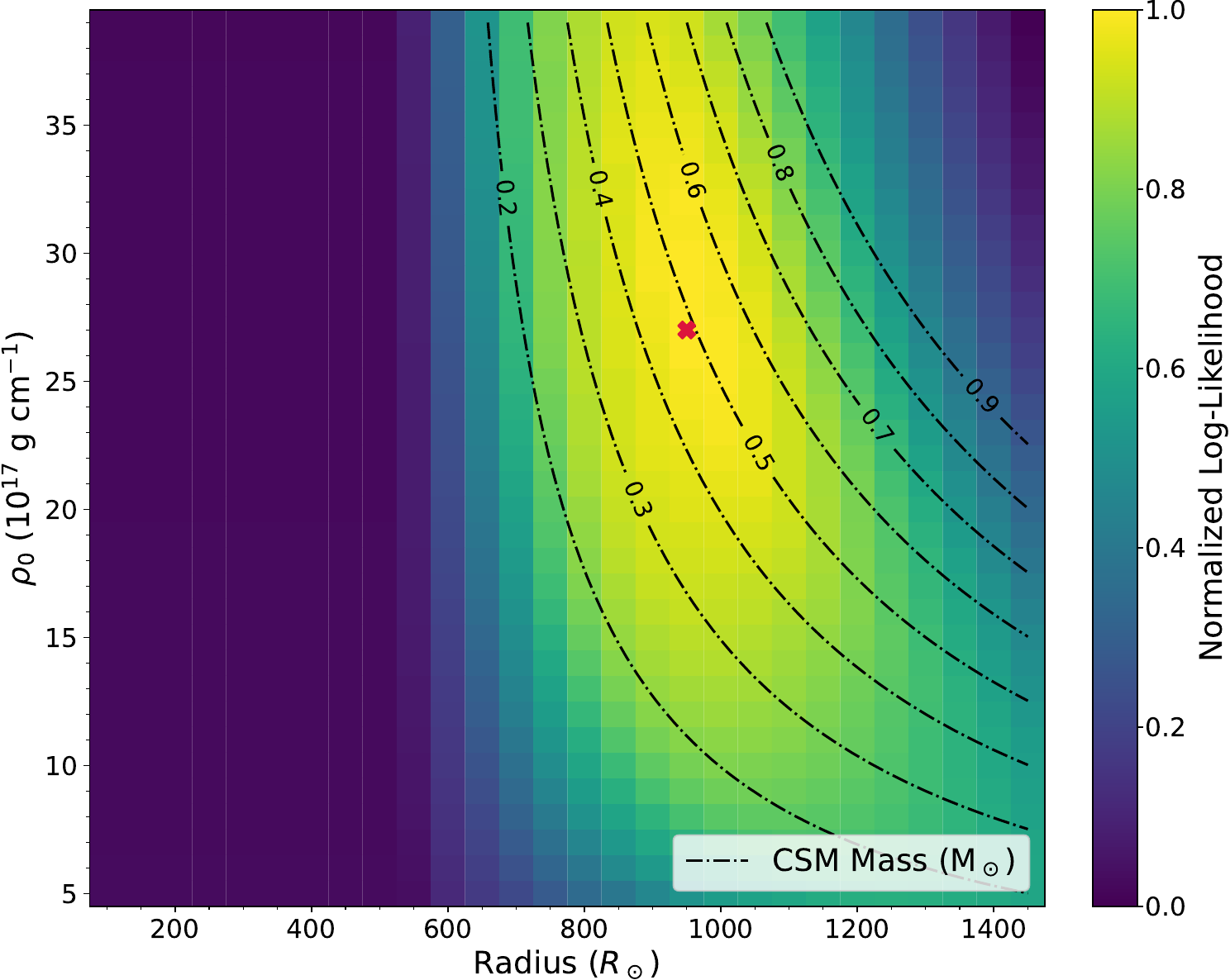}
    
    \end{minipage}
    
    \vspace{1em} 
    
    \includegraphics[width=0.5\textwidth]{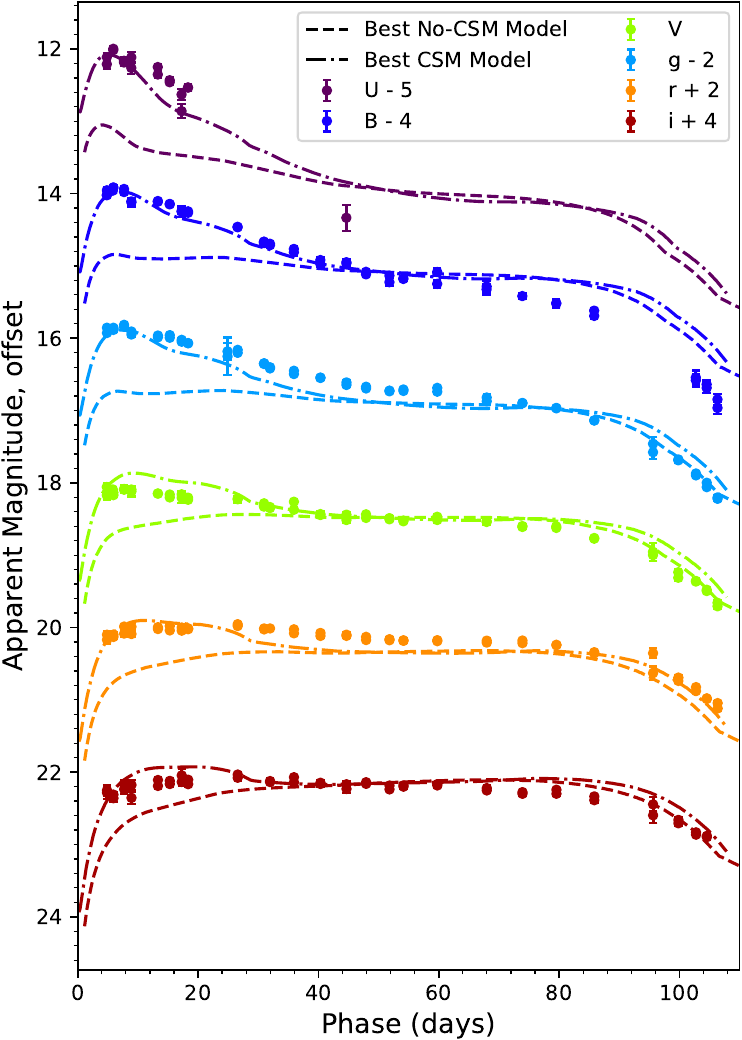}
    \caption{\emph{Left panel:} No-CSM \textsc{snec} model grid heatmap.  Lighter bins represent more likely models, i.e. better-fitting models.  The maximum occurs at $10.5M_\odot$ and $5.75\cdot10^{50}$ ergs, with degeneracy present in both parameters.  \emph{Right panel:} CSM model grid heatmap.  The maximum occurs at extent $950R_\odot$ and $\rho_0=2.7\cdot10^{18}\text{ g cm}^{-1}$, but with significant degeneracy in $\rho_0$.  The parameter space near the maximum at $\sim950R_\odot$ has similar likelihood, implying the presence of a CSM shell of radial extent $\sim950R_\odot$, but without well-constrained density.  Contours representing the total CSM mass are plotted above the heatmap.  \emph{Bottom panel:} $UBVgri$ light curves of all models, split by the presence of CSM.  The presence of CSM is required to reproduce the photometry during the first $30$ days but not afterwards.}
    \label{fig:snecsims}
\end{figure*}

 Since exploring the full four-dimensional parameter space would be prohibitively computationally expensive, we used the methods described in \cite{Morozova2017} and \cite{Morozova2018} to reduce the computational load.  After fixing the  $^{56}$Ni mass to our measured value of $0.047M_\odot$, we constructed a model grid exploring mass-energy parameter space with no CSM, comparing only to the light curve between the end of the early decline and $t_\text{PT}$.  While CSM primarily affects the early evolution, it can also influence the plateau height and duration in models with dense, extended CSM.  We subsequently fixed the mass and explosion energy to the most likely model's parameters, then explored the CSM density-extent parameter space, comparing to the full light curve pre-$t_\text{PT}$.  We define the hydrogen envelope mass $M_{H_\text{env}}$ to be the mass above the $20\%$ hydrogen mass fraction point $X\geq0.2$ as in \cite{Hiramatsu2021}.  The results of the model grids are depicted in Figure \ref{fig:snecsims}, and we summarize our most likely model parameters in Table \ref{tab:best_models}.

\par The most likely model without CSM fits the photometric evolution after the early decline, but overestimates the $i$-band flux somewhat uniformly across the plateau, while the most likely model with CSM better replicates the early decline in the $U$, $B$, and $g$ bands, but overestimates the $i$-band flux.  Overall, the \textsc{snec} models suggest the most likely scenario involves a $\sim10.5M_\odot$ progenitor with an explosion energy $\sim5.75\times10^{50}$ ergs enclosed in a CSM shell with an extent $\sim950R_\odot$ and $\rho_0\sim2.7\times10^{18} \text{ g cm}^{-1}$.  The suggested progenitor mass is consistent with the upper limit discussed in Section \ref{sec:nebular}.  Additionally, the CSM parameter space is highly degenerate, making it difficult to draw any robust conclusions about the CSM configuration of SN\,2019hnl other than it not being of both large density and radial extent.  The degeneracy in $\rho_0$ and extent is likely a result of multiple CSM configurations being able to produce the same total CSM mass, as \textsc{snec} models tend to be sensitive to total CSM mass in this region of parameter space \citep{SNEC}.

\subsubsection{Hydrodynamic Modeling: MESA+STELLA} \label{sec:mesa}

\par As discussed in Section \ref{sec:nebular}, SN\,2019hnl has a lower hydrogen content than models in its nebular phase, implying partial hydrogen envelope stripping during the progenitor star's evolution.  To explore the effects of envelope stripping on the light curve of SN\,2019hnl, we constructed a \textsc{mesa} \citep{Paxton2011, Paxton2013, Paxton2015, Paxton2018, Paxton2019, Jermyn2023} + \textsc{stella} \citep{Stella1, Stella2, Stella3} model grid.  We model mass loss with the ``Dutch" wind model for massive stars \citep{Dutch4, Dutch3, Dutch2, Dutch1}, varying the Dutch wind scaling factor $\eta$ to alter the extent of envelope stripping.  Greater $\eta$ correspond to smaller $M_{H_\text{env}}$ due to wind-driven mass loss.

\par We begin our simulations with nonrotating, pre-MS, $Z_\odot$ models.  We then evolve these models with \textsc{mesa} through the MS, post-MS, core collapse, explosion, and shock propagation nearly up to the surface of the star, at which point we cut the model at the optical depth $\tau=2/3$ and enclose the star in an $r^{-2}$ CSM density distribution.  We then transfer the model to \textsc{stella}, which simulates the subsequent breakout and photometric evolution.\footnote{This is in accordance with \textsc{mesa} test suite cases \texttt{12M\_pre\_ms\_to\_core\_collapse} and \texttt{ccsn\_IIp}, both in \textsc{mesa} version \texttt{24.08.1}.  See \cite{Paxton2018, Paxton2019} for a comprehensive description.}  Since \textsc{stella} model grids are frequency-dependent and do not presume pure LTE, \textsc{stella} is capable of simulating post-$t_\text{PT}$ photometric evolution.  Additionally, \textsc{stella} does not natively support photometry in the $gri$ bands.  Therefore, we compare with our $UBV$ photometry past the early decline for non-CSM models, and our full $UBV$ light curve for CSM models.

\par
\begin{figure*}
    \centering
    \begin{minipage}{0.48\textwidth}
        \includegraphics[width=\linewidth]{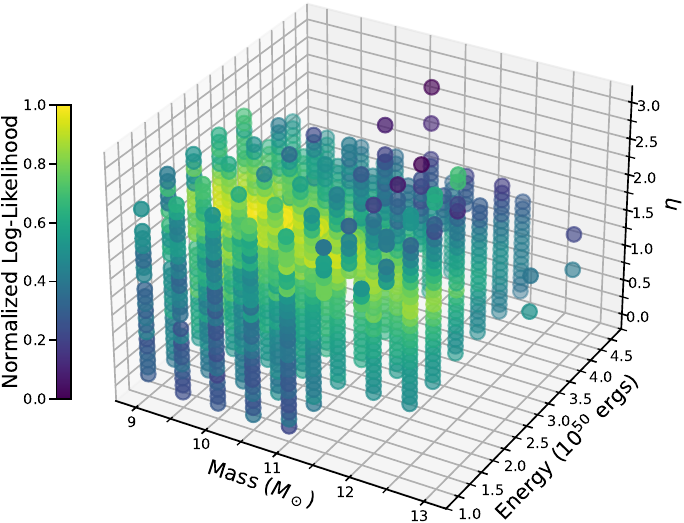}
    \end{minipage}\hfill
    \begin{minipage}{0.48\textwidth}
    \includegraphics[width=\linewidth]{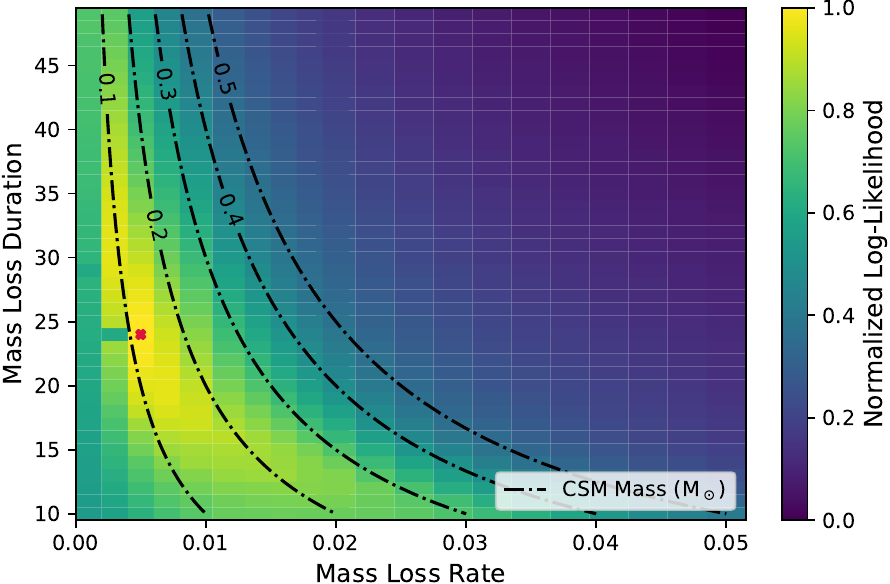}
    
    \end{minipage}
    
    \vspace{1em} 
    
    \includegraphics[width=0.5\textwidth]{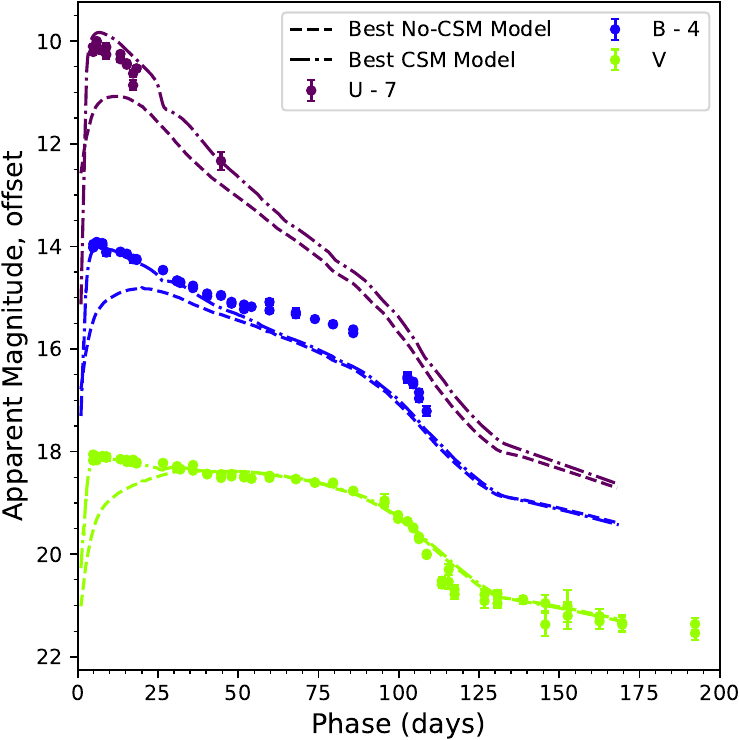}
    \caption{\emph{Left panel:} No-CSM \textsc{mesa} + \textsc{stella} model grid heatmap scatterplot, with lighter regions representing areas of greater likelihood.  The maximum occurs at $M_\text{ZAMS}=10.0M_\odot$, $E=2.4\cdot10^{50}$ ergs, $\eta=1.7$, and $M_{H_\text{env}}=4.3M_\odot$, within a locally likelier volume, introducing degeneracy.  \emph{Right panel:} CSM model grid heatmap.  The maximum occurs at $\dot{M}=5\cdot10^{-3}M_\odot\text{ yr}^{-1}$ and $t_\text{wind}=24 \text{ yr}$, with degeneracy in $t_\text{wind}$.  Contours representing the total CSM mass are plotted above the heatmap.  Similarly to the \textsc{snec} models, there appears to be degeneracy related to total CSM mass.  \emph{Bottom panel:} $UBV$ light curves of all models, split by the inclusion of CSM.  The best fit for each model set are shown in dashed lines.  The early behavior of the light curve is nonreplicable without CSM, which also raises the $B$-band brightness to observed levels during the early decline.  The plateau, fall therefrom, and $^{56}$Ni tail are all well-reproduced aside from SN\,2019hnl's bluer-than-typical plateau discussed in Section \ref{sec:reddening}.}
    \label{fig:mesastella}
\end{figure*}

 Using the same technique as in Section \ref{sec:snec}, we first created a model grid with $M_\text{ZAMS}$, explosion energy, and $\eta$ as free parameters, with no CSM added.  Since we used the \textsc{snec} results to broadly initialize our parameters, the models quickly converged to a maximum.  However, due to the higher computational cost of the \textsc{mesa}+\textsc{stella} models, we explored a narrower range of mass and energy.  After comparison and determination of the most likely model, we fixed $M_\text{ZAMS}$, explosion energy, and $\eta$, then introduced CSM by setting $\dot{M}$ and the mass loss duration $t_\text{wind}$ as free parameters, fixing wind velocity at $10\text{ km s}^{-1}$ \citep{WindVelocity}.  The results of our models are shown in Figure \ref{fig:mesastella} and the characteristics of the most likely model are tabulated in Table \ref{tab:best_models}.

\begin{table*}
    \centering
    \begin{tabular}{@{}lcccc@{}}
    \toprule
    \textbf{SNEC} \\
    \midrule
    \multicolumn{1}{c}{\textbf{Section}} & \textbf{Parameter} & \textbf{Variable} & \textbf{Value} & \textbf{Parameter Type}\\ 
    \midrule
    \multicolumn{1}{c}{\textbf{Progenitor}} & ZAMS mass & $M_\text{ZAMS}$ & $10.5M_\odot$ & Free, controlled \\
    & Explosion energy & $E$ & $5.75\cdot10^{50}$ ergs & Free, controlled \\
    & Hydrogen envelope mass & $M_{H_\text{env}}$ & $7.52M_\odot$ & Free, not controlled \\
    & Radius & $R$ & $542R_\odot$ & Free, not controlled \\
    \midrule
    \multicolumn{1}{c}{\textbf{CSM}} & CSM extent & $R_\text{CSM}$ & $950R_\odot$ & Free, controlled \\
    & Density scaling parameter & $\rho_0$ & $ 2.7\cdot10^{18}\text{ g cm}^{-1}$ & Free, controlled \\
    & Total CSM mass & $M_\text{CSM}$ & $0.48M_\odot$ & Derived \\
    \midrule
    \textbf{MESA+STELLA} \\
    \midrule
    \multicolumn{1}{c}{\textbf{Progenitor}} & ZAMS mass & $M_\text{ZAMS}$ & $10.0M_\odot$ & Free, controlled \\
    & Explosion energy & $E$ & $2.4\cdot10^{50}$ ergs & Free, controlled \\
    & Dutch wind scaling factor & $\eta$ & $1.7$ & Free, controlled \\
    & Hydrogen envelope mass & $M_{H_\text{env}}$ & $4.3M_\odot$ & Free, not controlled \\
    & Radius & $R$ & $1046R_\odot$ & Free, not controlled \\
    \midrule
    \multicolumn{1}{c}{\textbf{CSM}} & Mass loss duration & $t_\text{wind}$ & $24$ yr & Free, controlled \\
    & Mass loss rate & $\dot{M}$ & $5\cdot10^{-3}M_\odot \text{ yr}^{-1}$ & Free, controlled \\
    & Wind velocity & $v$ & $10\text{ km s}^{-1}$ & Fixed \\
    & Total CSM mass & $M_\text{CSM}$ & $0.12M_\odot$ & Derived\\
    \midrule
    \bottomrule
    \end{tabular}
    \caption{Most likely model parameters for our \textsc{snec} and \textsc{mesa}+\textsc{stella} model grids.}
    \label{tab:best_models}
\end{table*}

\par The most likely model devoid of CSM fits the photometric evolution after early decline well in the $V$ band, maintaining similar brightnesses and falling from plateau at the appropriate time and rate, however it underestimates the $B$ band brightness.  As anticipated, the model also fails to replicate the rise.  However, the most likely CSM model well-replicates the rise and brightens the $B$ band at early times, accurately fitting the early photometric evolution.  For these reasons, we infer that the models match well with observations.

\par The model grids suggest a low-mass progenitor $M_\text{ZAMS}\sim10M_\odot$ with $M_{H_\text{env}}\sim4.3M_\odot$ and a substantially lower explosion energy $\sim2.4\times10^{50}\text{ ergs}$ than suggested in Section \ref{sec:snec}.  This difference is possibly due to the lower stripping in the \cite{Sukhbold2016} models compared to the \textsc{mesa} models, causing a photometric difference for which greater energies and masses could compensate due to degeneracy.  The derived mass is consistent with both the modeling discussed in Section \ref{sec:snec} and the $\lesssim12M_\odot$ limit found in Section \ref{sec:nebular}.  Additionally, the lower mass estimate is consistent with the $12\pm1M_\odot$ mass estimate of the slightly brighter, photometrically similar SN\,1999em determined with pre-explosion photometric limits and stellar evolution tracks \citep{1999em_mass}.  As more massive progenitors tend to result in more luminous plateaus, our $\sim10M_\odot$ estimate is qualitatively consistent with SN\,1999em's $12\pm1M_\odot$ estimate.  Furthermore, the lower energy estimate is consistent with the subtypical \ion{Fe}{2} velocity found in Section \ref{sec:spec_evo}.

\par As anticipated, $M_{H_\text{env}}$ is lighter than more typical SN\,IIP progenitors \citep[$M_{H_\text{env}} \gtrsim 4.5M_\odot$;][]{Hiramatsu2021} but still substantial, while $R$ places the progenitor within typical RSG sizes \citep{RSG_Radius} in alignment with ordinary SNe\,IIP progenitors \citep{RSG_Prog}.  The most likely CSM configuration has $t_\text{wind}\sim24\text{ yr}$ and $\dot{M}\sim5\times10^{-3} M_\odot \text{ yr}^{-1}$.  The high $\dot{M}$ prior to explosion is consistent with that estimated in other SNe\,IIP \citep{Moriya2011, Morozova2018}, but is also near the upper limit of mass loss rates for stellar winds of $\sim\mathcal{O}(10^{-3}) M_\odot \text{ yr}^{-1}$ \citep{winds_are_real}.  Additionally, the short mass loss duration is inconsistent with the longer timescales anticipated of stellar winds.  Given that the high mass loss rates and short timescales derived here can be produced by binary systems \citep{binary1, binary2, binary3, binary4}, we tentatively interpret these results as indicative of possible binary interaction.  Such a result would also be consistent with the extent of the stripping experienced by the progenitor.

\par
\begin{figure}
    \centering
    \includegraphics[width=\linewidth]{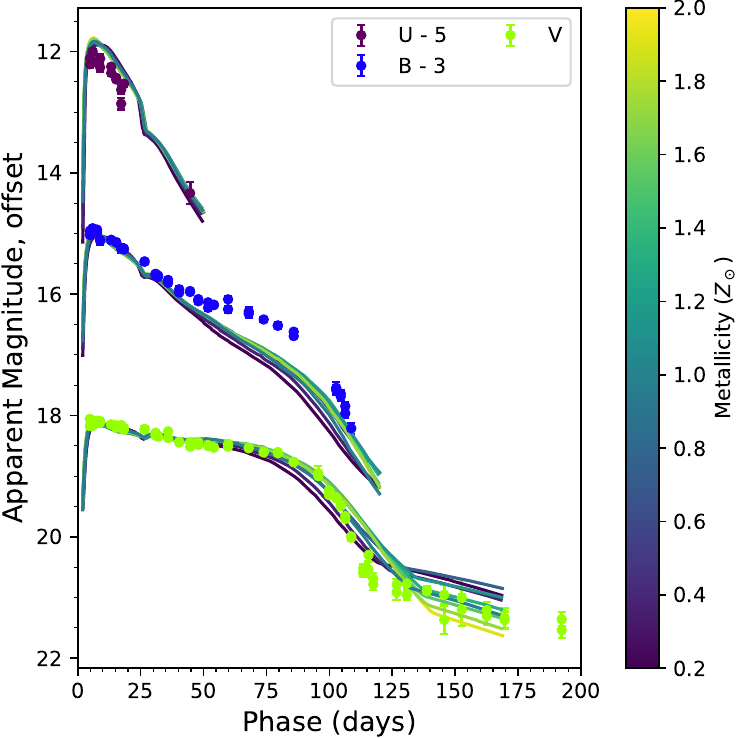}
    \caption{$UBV$ light curves of the models with variable metallicity.  The early behavior of the light curve is essentially invariant with metallicity, while minor variance begins to occur near to and after the fall from plateau.}
    \label{fig:z_lightcurves}
\end{figure}

 Given the uncertainty of the metallicity discussed in Section \ref{sec:spec_evo}, we ran a set of models with parameters identical to those of our best model (Table \ref{tab:best_models}) but with metallicities varying from $0.2Z_\odot$ to $2Z_\odot$ (Figure \ref{fig:z_lightcurves}) to determine if metallicity has significant effects on our results.  The photometric evolution remains relatively unchanged as metallicity varies, only beginning to differ near the end of plateau and onwards.  Given the relative insensitivity of the light curve to metallicity, we believe the uncertainty in metallicity does not likely impact our modeling results significantly.

\section{Conclusion} \label{sec:conclusion}

\par In this paper, we present and analyze our photometric and spectroscopic observations of the Type IIP SN\,2019hnl.  SN\,2019hnl was discovered by the ATLAS survey within $\sim$1 day of the explosion, was followed photometrically until $\sim200$ days post-explosion, and spectroscopically between 5 and 428 days post-explosion.  We used the photometric and spectroscopic data collected to reach the following summarized results:

\begin{enumerate}
    \item SN\,2019hnl is a SN\,IIP of typical luminosity with $M_V = -16.7\pm0.1$ mag, a rise time $\sim6$ days, and a characteristic $t_\text{PT} \sim 107.4\pm0.4$ days.  The slope during the initial fall from maximum onto the plateau is regular for SNe\,IIP ($S_{50}\sim0.0086\pm0.0006 \text{ mag (50 days)}^{-1}$).  SN\,2019hnl is solidly located within SNe\,IIP photometric parameter spaces.

    \item From constructing a pseudobolometric light curve of SN\,2019hnl and analyzing its tail, we estimate a $^{56}$Ni yield $\sim 0.047 \pm 0.007M_\odot$ from the explosion, which is typical for SNe\,II of its $M_V$ range.  Complete trapping was assumed and is consistent with the tail decline rate.

    \item The plateau spectra of SN\,2019hnl show Fe $\lambda5018$ pEWs similar to that of metal-poor models at early times but migrate to more metal-rich models at later times \citep{Z_Comp}.  While the \ion{Fe}{2} lines are weak in our spectra, their evolution does not match well with a single model, possibly due to a faster temperature drop.  We conclude that the evolution is likely degenerate with other explosion properties, making definite estimates problematic.

    \item We compared our $t+428$d nebular spectrum to model spectra described in \cite{JerkstrandModels} and estimate, based on the relative intensity of the \ion{O}{1} $\lambda\lambda6300,6364$ doublet, an upper limit of $\sim12M_\odot$ for the ZAMS mass of SN\,2019hnl's progenitor.  Furthermore, we attribute the relatively weak H$\alpha$ line to partial stripping of the progenitor.

    \item We constructed model grids with the \textsc{snec} and \textsc{mesa}+\textsc{stella} codes.  We find that both the non-stripped \textsc{snec} models and the partially stripped \textsc{mesa}+\textsc{stella} models are consistent with our upper limit on mass from nebular spectroscopy.  The \textsc{snec} models also suggest a relatively dense shell of CSM.  In alignment with the estimated mass of the photometrically similar SN\,1999em, we therefore estimate a progenitor ZAMS mass of $\sim10M_\odot$ and a pre-explosion mass of $\sim9M_\odot$ from our \textsc{snec} and \textsc{mesa}+\textsc{stella} modeling.  Furthermore, our \textsc{mesa}+\textsc{stella} CSM modeling suggests a dense shell of CSM around the progenitor resulting from a high mass loss rate $\dot{M}\sim5\cdot10^{-3} M_\odot \text{ yr}^{-1}$ for $\sim24\text{ yr}$ preceding explosion, typical of partially stripped SNe \citep{Hiramatsu2021}.  We find that despite SN\,2019hnl's standard photometric evolution, its progenitor likely experienced partial stripping during its evolution and underwent significant mass loss preceding explosion, possibly as a result of binary interaction.
\end{enumerate}

\par In conclusion, we find that SN\,2019hnl is a typical SN\,IIP with an ordinary photometric evolution that resulted from the explosion of a partially stripped progenitor.  The increasingly large sample of partially stripped, markedly typical SNe\,II is suggestive of a general trend towards stripping being more common than previously anticipated.  Further investigation into spectral modeling of partially stripped SNe\,II could break degeneracies in photometric modeling and shed light on their explosion physics.

\section*{Acknowledgments}

% ref
\par We thank the anonymous referee for their valuable and constructive consideration of our work.

% DLT40
\par Research by the DLT40 survey is supported by National Science Foundation (NSF) grant AST-2407565.

% Univ. Arizona
\par Time-domain research by the University of Arizona team and D.J.S. is supported by National Science Foundation (NSF) grants 2108032, 2308181, 2407566, and 2432036 and the Heising-Simons Foundation under grant \#2020-1864. 

\par K.A.B. is supported through the LSST-DA Catalyst Fellowship project; this publication was thus made possible through the support of Grant 62192 from the John Templeton Foundation to LSST-DA.  The opinions expressed in this publication do not necessarily reflect the views of LSST-DA or the John Templeton Foundation.  
% LCO
\par This work makes use of observations from the Las Cumbres Observatory network.
The Las Cumbres Observatory team is supported by NSF grants AST-1911225 and AST-1911151.

\par Supernova research at Rutgers University is supported in part by NSF award 2407567 to S.W.J.

\par Some of the data presented herein were obtained at Keck Observatory, which is a private 501(c)3 non-profit organization operated as a scientific partnership among the California Institute of Technology, the University of California, and the National Aeronautics and Space Administration.  The Observatory was made possible by the generous financial support of the W. M. Keck Foundation.  The authors wish to recognize and acknowledge the very significant cultural role and reverence that the summit of Maunakea has always had within the indigenous Hawaiian community.  We are most fortunate to have the opportunity to conduct observations from this mountain.

\par This research made use of the NASA/IPAC Extragalactic Database (NED; 10.26132/NED1), which is funded by the National Aeronautics and Space Administration and operated by the California Institute of Technology.

\par This research has made use of the VizieR catalogue access tool, CDS,
Strasbourg, France \citep{10.26093/cds/vizier}. The original description 
of the VizieR service was published in \citet{vizier2000}.

\vspace{5mm}
\facilities{LCOGT, ATLAS, Keck:I (LRIS)}

\software{Astropy \citep{Astropy1, Astropy2, Astropy3}, NumPy \citep{NumPy}, Matplotlib \citep{Matplotlib}, Pandas \citep{Pandas}, SciPy \citep{SciPy}, emcee \citep{emcee}, \textsc{lcogtsnpipe} \citep{Valenti2016}, \textsc{snec} \citep{SNEC}, \textsc{mesa} \nocite{242}\citep{Paxton2011, Paxton2013, Paxton2015, Paxton2018, Paxton2019, Jermyn2023}, \textsc{stella} \citep{Stella1, Stella2, Stella3}, Siril \citep{Siril}}

\appendix

\setcounter{table}{0}
\renewcommand{\thetable}{A\arabic{table}}

\section{Expanding Photosphere Method} \label{appendix:epm}

\par We applied the expanding photosphere method (EPM) to SN\,2019hnl in an attempt to determine an independent distance estimate.  EPM assumes that the photosphere is a dilute blackbody and is expanding spherically and unrestricted.  With these assumptions, we can relate distance to photospheric velocity $v_\text{phot}$, angular size $\theta$, and time since explosion $t-t_0$ with Equation \ref{eqn:epm1}, and minimize $\epsilon$ in Equation \ref{eqn:epm2} to find $\theta$ and the color temperature $T_c$

\begin{align}
    D &= (t-t_0)\frac{v_\text{phot}}{\theta}
    \label{eqn:epm1}\\
    \epsilon &= \sum_{\nu\in S} \{m_\nu + 5\log\big(\theta\xi(T_c)\big) - A_\nu - b_\nu(T_c)\}^2
    \label{eqn:epm2}
\end{align}

where $m_\nu$ is the apparent magnitude from filter $\nu$ in filter set $S$, $\xi$ the dilution factor, and $b_\nu$ the synthetic magnitude.  Due to the absence of sufficient spectra during early decline containing the \ion{Fe}{2} $\lambda5169$ line, we used the $v_\text{H$\beta$}\leftrightarrow v_\text{\ion{Fe}{2}}$ transformation of \cite{HbetaConv} to estimate \ion{Fe}{2} velocities from Gaussian fits to the H$\beta$ P Cygni profile absorption minima.  The H$\beta$ and extrapolated \ion{Fe}{2} velocities are summarized in Table \ref{tab:v_feii}.

\begin{table}[h]
    \centering
    \caption{Measured H$\beta$ emission velocities and the transformed \ion{Fe}{2} velocities from spectral continua.}
    \begin{tabular}{c|c|c}
         Phase (days) & $v_{\text{H}\beta}$ (km s$^{-1}$) & $v_{\text{\ion{Fe}{2}}}$ (km s$^{-1}$) \\
         \hline
         $5.60$ & $8121\pm97.6$ & $6822\pm82.0$ \\
         $15.64$ & $7822\pm39.9$ & $6570\pm33.5$ \\
         $20.61$ & $7590\pm76.3$ & $6376\pm64.1$ \\
         $30.63$ & $6249\pm105.1$ & $5249\pm88.3$ \\
    \end{tabular}
    \label{tab:v_feii}
\end{table}

\par To find $m_\nu$, we linearly interpolated our photometric data to the requisite times.  Both $\xi$ and $b_\nu$ can be expressed in terms of $T_c$ \citep{epm1, epm2}.  To minimize the equation, we used a Markov-chain Monte Carlo (MCMC) sampler with the $BV$, $BVI$, and $VI$ filter combinations.  Since we have no native $I$-band photometry, we synthesized points by transforming our $r$ and $i$ photometry in accordance with \cite{bandconv}.  The results are shown in Table \ref{tab:epmres}.  The \cite{bandconv} transformations are optimized for stellar spectra, which resemble a Planck distribution significantly more closely than our SN spectra.  As a result, this difference may have introduced systematic error in the synthesized $I$ band photometry.

\begin{table}[h]
    \centering
    \caption{Synthesized EPM distances, ordered by bandpass combination.  The distances are substantially variant, possibly a result of sparse spectral data and/or greater uncertainty from the H$\beta\rightarrow$\ion{Fe}{2} transformation.}
    \begin{tabular}{c|c}
         Bandpasses & Distance (Mpc)\\
         \hline
         $BV$ & $121.6^{+4.6}_{-4.3}$\\
         $BVI$ & $94.8^{+2.2}_{-2.2}$\\
         $VI$ & $78.0^{+2.7}_{-2.5}$
    \end{tabular}
    \label{tab:epmres}
\end{table}

\par The distance measurements vary significantly with bandpass combination.  Though the mean distance is similar to the redshift-derived distance, the high variance is possibly due to the sparse spectral and photometric data covering an insufficiently large temporal window to yield reliable results, combined with the increased uncertainty due to transforming H$\beta$ velocities to \ion{Fe}{2} velocities.  We therefore adopt the Hubble flow distance discussed in Section \ref{sec:distance}.

\bibliography{bibliography}{}
\bibliographystyle{aasjournal}

\end{document}